\documentclass[a4paper,twocolumn,11pt,unpublished]{quantumarticle}
\pdfoutput=1
\usepackage[utf8]{inputenc}
\usepackage[english]{babel}
\usepackage[T1]{fontenc}
\usepackage{amsmath}
\usepackage{hyperref}

\usepackage{graphicx}
\usepackage{subfigure}
\usepackage{epstopdf}
\usepackage{graphicx}

\usepackage{amsfonts}
\usepackage{amssymb, color, bm}

\usepackage{currvita}

\usepackage[noend]{algpseudocode}
\usepackage{algorithm}
\usepackage{setspace}

\usepackage{physics}

\usepackage{multicol}
\usepackage{lipsum}
\usepackage{mwe}

\usepackage{tikz}
\usepackage{lipsum}

\begin{document}

\title{Quantum Integrated Sensing and Computation with Indefinite Causal Order}
%\title{Self-learning, Quantum Sensing and Quantum Error Mitigation via Time Reversal Symmetry Breaking}
%\title{Quantum Time Reversal Symmetry Breaking: Applications to Self-learning, Parameter Estimation, and Error Detection}

\author{Ivana Nikoloska}
\affiliation{Signal Processing Systems Group,  Department of Electrical Engineering, Eindhoven University of Technology, Eindhoven, 5612 AP, The Netherlands}
\email {i.nikoloska@tue.nl. }
%\thanks{The work of I. Nikoloska has been supported by Quantum Delta NL, grant number ... }

\maketitle

\begin{abstract}
Quantum operations with indefinite causal order (ICO) represent a framework in quantum information processing where the relative order between two events can be indefinite. In this paper, we investigate whether sensing and computation, two canonical tasks in quantum information processing, can be carried out within the ICO framework. We propose a scheme for integrated sensing and computation that uses the same quantum state for both tasks. 
The quantum state is represented as an agent that performs state observation and learns a function of the state to make predictions via a parametric model.
Under an ICO operation, the agent experiences a superposition of orders, one in which it performs state observation and then executes the required computation steps, and another in which the agent carries out the computation first and then performs state observation. This is distinct from prevailing information processing and  machine intelligence paradigms where information acquisition and learning follow a strict causal order, with the former always preceding the latter. We provide experimental results and we show that the proposed scheme can achieve small training and testing losses on a representative task in magnetic navigation. 

\end{abstract}

\section{Introduction }
\subsection{Context and Motivation} 
Can an agent process information before acquiring it? Surely, information processing implies strict causal order: the agent obtains information and only then can learn from it and act accordingly. This seems to describe both everyday experience for biological agents, as well as the basis of  seminal works in information processing for artificial ones \cite{sutton1998reinforcement}. For example, Weaver and Shannon separate this problem into three hierarchical levels: The technical level is concerned with information acquisition, e.g., via a wireless network, or on-device sensors; The  semantic level is concerned with extracting meaning from the acquired information, e.g., via a machine learning model, and the effectiveness level is concerned with acting upon the extracted meaning. Each level is dependent on the lower ones, and one can only execute tasks in a linear order, e.g., meaning can be extracted only if information has already been acquired \cite{weaver1953recent}.

\begin{figure*}
    \centering
    \includegraphics[width = 0.95\textwidth]{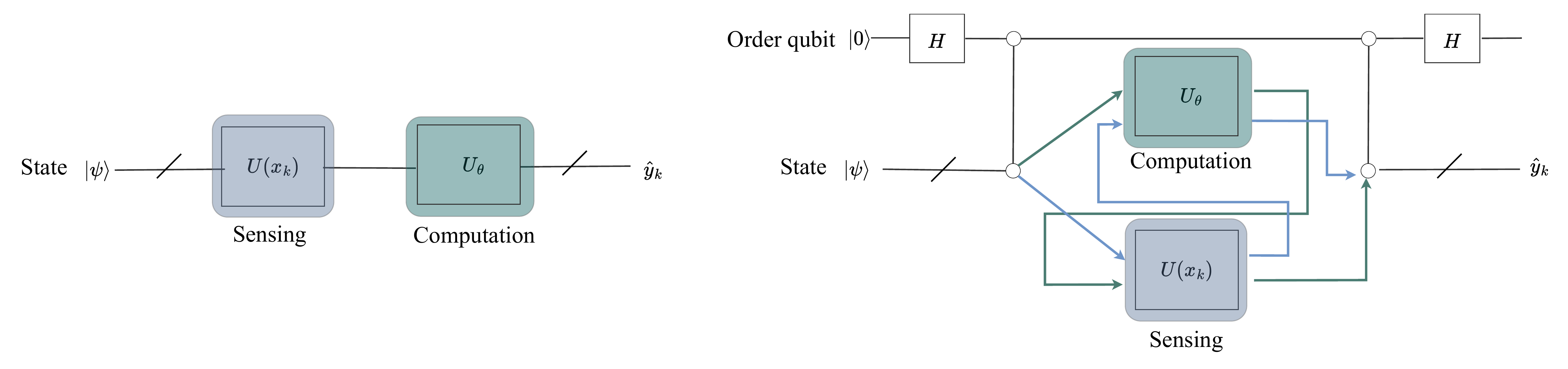}
    \caption{Quantum integrated sensing and computation with linear causal order  (left) and  ICO (right). In the proposed framework, sensing and computation are carried out using the same quantum state $\ket{\psi}$. Under ICO, the agent effectively experiences a superposition of orders, one in which it observes the state and then computes, and another where the agent carries out the computation and then observes the state.} 
    \label{fig:sensing_scheme_indirect} 
\end{figure*}

However, physical systems obey the laws of quantum mechanics, and quantum mechanics  is compatible with scenarios where the relative order between two events can be indefinite \cite{chiribella2009beyond}. This has enabled a new framework in quantum information processing named quantum operations with indefinite causal order (ICO). These ideas have been motivated by the grand challenge of uniting quantum mechanics with gravity \cite{hardy2007towards}, or to characterize quantum theory in terms of information processing \cite{brassard2005information, fuchs2003quantum}. In the ICO framework, the order of processes becomes entangled with a quantum degree of freedom. This type of entanglement, called causal non-separability, represents a new resource, which provides advantages in a variety of information processing tasks. Examples include quantum communication \cite{zhao2025quantum, guerin2016exponential}, quantum metrology \cite{wang2026quantum, yin2023experimental}, and inversion of unknown dynamics \cite{quintino2019reversing, quintino2019probabilistic}. In all these tasks, quantum operations with ICO were shown to outperform all possible quantum operations with definite ordering. Although the theoretical possibility of ICO represents a remarkable implication of the quantum framework, examining its information-theoretic consequences can actually provide deeper insight into quantum mechanics itself.

\subsection{Main Contributions} 
Motivated by these ideas, in this paper we investigate information acquisition and machine learning with ICO. To this end, as shown in Fig.~\ref{fig:sensing_scheme_indirect} (left), we first develop a framework for integrated sensing and computation in which both tasks are carried out using the same quantum state. In it, the quantum state is represented as an agent that performs state observation and learns a function of the state in order to make predictions using a parametric model. The ICO operation is realised using a quantum SWITCH that combines two operations in a quantum superposition of two alternative orders. Using the quantum SWITCH, Fig.~\ref{fig:sensing_scheme_indirect} (right), the agent effectively experiences a superposition of orders, one in which it observes the state and then computes, and another in which the agent carries out the computation and then observes the state. Whilst this investigation is primarily conceptual, we show that it also has engineering implications by providing extensive experiments on a real-world task in magnetic based navigation in robotics. 

\noindent \textbf{Organization:} The remainder of the paper is organized as follows. Sec.~II formulates quantum integrated sensing and computation. In Sec.~III we extend quantum integrated sensing and computation to include ICO. In Sec.~IV we provide experiments testing the performance of the proposed scheme, detailing the setup, benchmarks, and results. Sec.~V offers a discussion and concludes the paper.

\section{Quantum Integrated Sensing and Computation}\label{Sec:ICO}
In this section, we first provide a novel framework for quantum integrated sensing and computation. As shown in Fig.~\ref{fig:sensing_scheme_indirect} (left), in this framework, sensing and computation are carried out using a shared quantum state that performs state observation as a sensing probe and then computes a function of the state. Later, we extend the framework to include ICO of sensing and computation.

%\subsection{Quantum Integrated Sensing and Computation with Strict Causal Order}\label{Sec:causal}
Formally, we consider a $N$-qubit quantum state $\ket{\psi}$ representing an agent
that observes a state $x$ and then makes predictions $y$ based on $x$. For example, the agent can sense the magnetic field and then use the obtained information to predict its headings and navigate a terrain. We assume that $y = f(x)$ where the function $f(\cdot)$ is unknown and is approximated by a parametric model $U_{\theta}$, parametrised by variational parameters $\theta$. To this end, we assume a supervised learning setting where we have access to a dataset $\mathcal{D} = \{x_k,y_k\}_{k=1}^K$ of $K$ states $x_k$ and corresponding targets $y_k$. We note that the framework can also be extended to unsupervised learning, or reinforcement learning settings \cite{goodfellow2016deep}.

The quantum state $\ket{\psi}$ is assumed to be prepared by a (parametrised) quantum circuit. 
%The quantum circuit can be parametric in order to optimise the state, or one can prepare standard metrologically useful quantum states, such as GHZ states. 
Observing the state $x$ results in a perturbed quantum state as
\begin{align}\label{interact}
    \ket{\psi(x_k)} = U (x_k) \ket{\psi},
\end{align}
where $U (x_k) \ket{\psi}$ denotes the application of state-dependent unitary $U (x_k)$ to state $\ket{\psi}$. Note that, in practice, one has no control over $U (x_k)$ and how $x_k$ is imprinted on the state $\ket{\psi}$. 

The perturbed state $\ket{\psi(x_k)}$ is then used to make predictions $\hat{y_k}$ approximating the function $f(x_k)$ i.e., $\hat{y}_k \approx f(x_k)$. To make a prediction, we use a parametric model $U_{\theta}$ that transforms $\ket{\psi(x_k)}$ into an output state as 
\begin{align}\label{compute}
    \ket{\psi_{\theta}(x_k)} = U_{\theta} \ket{\psi(x_k)}.
\end{align}
%Unlike standard VQS frameworks aiming to estimate the parameter $x$ and obtain $\hat{x}$, we note that here we are interested in $\hat{y} \approx f(x)$
The output state $\ket{\psi_{\theta}(x_k)}$ is then measured which finally results in $\hat{y}_k$ as
\begin{align}\label{pred}
    \hat{y}_k = \langle O \rangle_{{\psi_{\theta}(x_k)}}.
\end{align}
This setup differs from standard quantum sensing frameworks that aim to obtain an estimate $\hat{x_k}$ of the parameter $x_k$ \cite{meyer2021variational, maclellan2024end,nikoloska2025dynamic, nikoloska2025adaptivebayesiansingleshotquantum, nikoloska2025variational}.
The goal here is to optimise the parameters of the model $U_{\theta}$ and minimise a loss function $\mathcal{L}_{\theta}(x_k)$ between the prediction $\hat{y}_k$ and the true target $y_k$. For example, the loss can be defined as the mean-squared loss, 
\begin{align}
    \mathcal{L}_{\theta}(x_k) = (y_k-\hat{y}_k)^2,
\end{align}
or the absolute loss
\begin{align}
    \mathcal{L}_{\theta}(x_k) = |y_k-\hat{y}_k|,
\end{align}
where $\hat{y}_k \approx f(x_k)$. Formally, we aim to solve
\begin{align}\label{prob_l}
    \underset{\theta}{\text{arg min}} \,\,\, \mathcal{L}_{\theta}(x_k),
\end{align}
where the optimisation is done over the variational parameters $\theta$. This problem is solved via gradient descent whereby the parameters are updated as
\begin{align}\label{grad}
    \theta \leftarrow \theta - \eta \frac{\partial \mathcal{L}_{\theta}(x_k)}{\partial\theta},
\end{align}
where $\eta$ denotes a learning rate and where the gradients can be computed via parameter-shift rules on actual quantum hardware, or standard backpropagation on a simulator \cite{nikoloska2026machine}.
Training can be carried out during an initial training phase in which the states and targets are known by design. 
%For example, in MRI the magnetic field is controlled by applying a known, periodic, sinusoidal modulation, with a known phase \cite{feiner1980nmr}. 
After training in such controlled conditions, assuming transferability, one could then use the same variational parameters when the states and corresponding targets are unknown. 
%The complete procedure is given in pseudo code in Algorithm 1.

\section{Quantum Integrated Sensing and Computation with ICO}\label{Sec:ICO}
The proposed integrated sensing and computation scheme can be extended to include ICO. To this end, we use a quantum SWITCH, a paradigmatic example of a higher-order operation that combines two input gates in a quantum superposition of two alternative orders. In particular, as shown in Fig.~\ref{fig:sensing_scheme_indirect} (right), we assume that the system is coupled to an order-qubit initialised as 
\begin{align}\label{control}
    \ket{+}_T = \frac{1}{\sqrt{2}} (\ket{0}_T + \ket{1}_T),
\end{align}
which controls the evolution of the system: If $T=0$, the system evolves under  $U_{\theta} U(x_k)$; if $T=1$, the system evolves under  $U (x_k) U_{\theta}$. We note that we cannot perform state observation with the order qubit, and we cannot use it for computation. The order qubit is part of the definition of the system \cite{chiribella2021indefinite}. 
%As noted in \cite{salek2018quantum}, in the gravitational realization of the quantum SWITCH, the order qubit would correspond to two alternative configurations of the masses in the universe, whereby this operation could even physically arise. 

Using the SWITCH, the complete system evolves under 
%\begin{align}\label{time_Q}
%    \ket{\Psi_{\theta}(x)} &= \frac{1}{\sqrt{T}} (\ket{0}_T\bra{0} \otimes U_{\theta} U(x) \ket{\psi} \nonumber\\
%    &+ \ket{1}_T\bra{1} \otimes U(x) U_{\theta}\ket{\psi}).
%\end{align}
\begin{align}\label{time_Q}
    U_{\theta}(x_k) &= \ket{0}_T\bra{0} \otimes U_{\theta} U(x_k)  \nonumber\\
    &+ \ket{1}_T\bra{1} \otimes U(x_k) U_{\theta}.
\end{align}
In words, the agent experiences a superposition of orders of $U(x_k)$ and $U_{\theta}$, one where it observes the state and then processes this information, and another where the agent carries out the computation and then performs state observation. This notion challenges the usual idea that time flows in a single, definite direction and that cause (i.e., information acquisition) always precedes effect (i.e., learning) in quantum information processing. 

The output state of the computation qubits $\ket{\Psi_{\theta}(x_k)}$ is then measured in order to obtain a prediction $\hat{y}$ as
\begin{align}\label{pred_ICO}
    \hat{y}_k = \langle O \rangle_{{\Psi_{\theta}(x_k)}}.
\end{align}
Similarly to the previous section, the goal is to optimise the parameters of the model $U_{\theta}$ in order to minimise a loss function $\mathcal{L}_{\theta}(x_k)$ between the prediction $\hat{y}_k$ and the true target $y_k$ as
\begin{align}\label{prob_l1}
    \underset{\theta}{\text{arg min}} \,\,\, \mathcal{L}_{\theta}(x_k).
\end{align}
This is done by updating the variational parameters $\theta$ via gradient descent as in \eqref{grad} where the gradients can be computed via parameter-shift rules on actual quantum hardware, or standard backpropagation on a simulator \cite{nikoloska2026machine}. 

\begin{figure*}[htbp]
    \centering
    \subfigure{\includegraphics[width=0.41\textwidth]{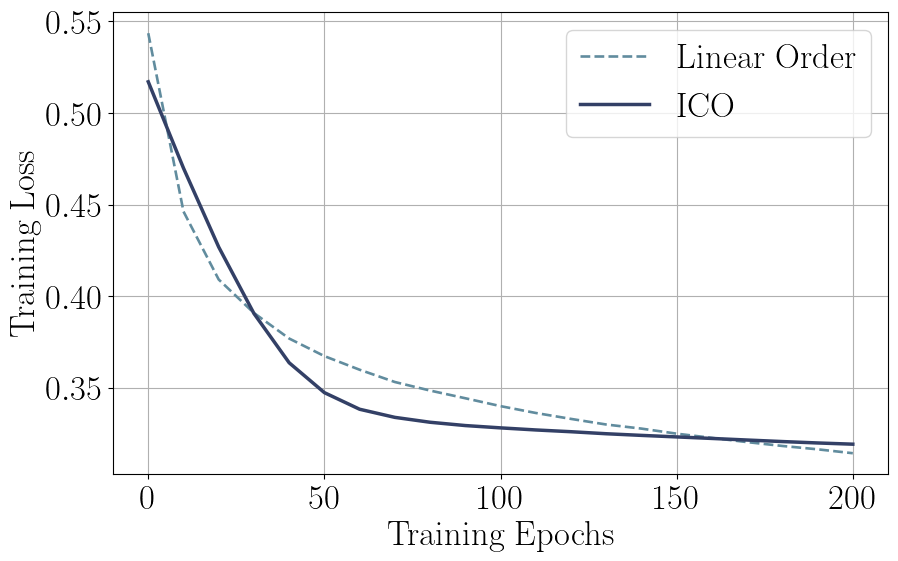}}
    \subfigure{\includegraphics[width=0.41\textwidth]{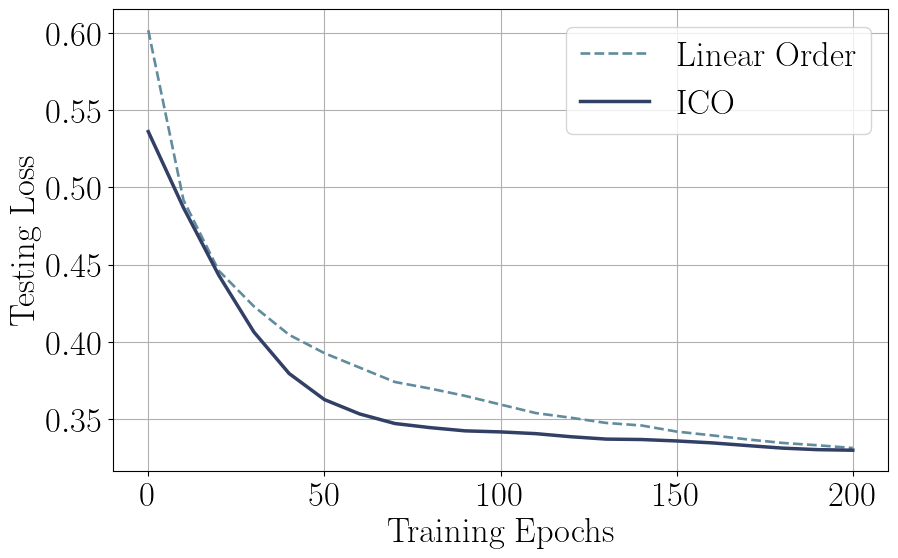}}
    \caption{Accrued training loss (left) and testing loss (right) over training epochs. We use a hardware-efficient model $U_{\theta}$ comprised of single qubit parametric gates and entangling CNOT gates with cyclical connectivity. The ansatz is comprised of $15$ layers and $N=3$ qubits. We use an additional qubit as an order qubit in the ICO scheme.}
    
    \label{fig:entropy_Tr}
\end{figure*}

%This notion challenges the usual idea that time flows in a single, definite direction and that cause always precedes effect and it is meant to represent a theoretical exercise. ...
%However, it also has a more grounded, engineering interpretation. In the reverse order, the unitary $U_{\theta}$ can also be seen as enhancing the probe and its sensitivity to the parameter $x$ as it is common in VQS, not just means to extract a prediction. However, unlike in VQS the only goal here is to carry out computation and learn the function of $x$ by minimising the adopted loss function $\mathcal{L}_{\theta}(x)$ in \eqref{prob_l}.

 \section{Experiments}
In this section, we provide experimental results to validate the proposed ICO scheme. The code for the experiments will be available on Github.

\subsection{Task}
%We consider two tasks:
%\subsubsection{Magnetometry} 
We consider the task of navigation, where an agent senses the magnetic field and learns to compute headings in order to navigate a terrain \cite{akai2014development, ataka2022magnetic}. We specifically use this task since magnetometry is considered to be a practical application of quantum sensing \cite{aiello2013composite, maayani2019distributed}. 

In particular, the agent senses the magnetic vector $x_k = \{B_{x_k}, B_{y_k}, B_{z_k}\}$, with $B_{x_k}, B_{y_k}, B_{z_k}$ indicating the strength in the $x, y, z$ directions, respectively. The magnetic vector is imprinted onto the $N=3$ qubit state $\ket{\psi}$ as 
\begin{align}
    U(x_k)&= ({R_{X}}(B_{x_k}) \otimes I \otimes I) \times \nonumber\\
    & \times (I \otimes {R_Y}(B_{y_k}) \otimes I ) \times ( I \otimes I \otimes {R_Z}(B_{z_k})).
\end{align}
where ${R_X}, {R_Y}, {R_Z}$ represent the Pauli $X, Y, Z$ rotation gates, respectively. The strength in each direction $B_{x_k}, B_{y_k}, B_{z_k}$ admits typical values for the field of the Earth at the surface. Specifically, $B_{x_k} \in (-43000, 43000)$nT, $B_{y_k} \in (-18000, 17000)$nT, and $B_{z_k} \in (-67000, 62000)$nT. The values are uniformly sampled from these intervals.

The agent's goal is to learn to predict the headings $y_k = f(x_k)$ which represent the direction it is facing at a moment, measured in degrees relative to a reference (here, north).
%thereby telling the orientation.
%, not necessarily the exact path it is traveling. 
For example, a heading of $25^o$ means that the agent is pointing $25$ degrees clockwise from north. The headings are calculated as
\begin{align}
    f(x_k) = 2 \arctan(B_{x_k}, B_{y_k}) 
\end{align}
and they are normalised to degrees. The predictions $\hat{y} \approx f(x_k)$ of the heading $y_k$ are learned by optimising the variational parameters $\theta$ of the parametric model $U_{\theta}$. This is done by minimising the mean squared loss function 
\begin{align}\label{mse}
    \mathcal{L}_{\theta}(x_k) = (y_k - \hat{y}_k)^2,
\end{align}
which we do via gradient descent using \eqref{grad}.

\subsection{Architecture and Hyperparameters}\label{sec:arch}
The quantum circuit used for learning the predictions is a hardware-efficient model comprised of single qubit parametric gates and entangling CNOT gates \cite{nikoloska2026machine}. The CNOT gates follow cyclical connectivity. The ansatz is comprised of $15$ layers and $N=3$ qubits. We use an additional qubit as an order qubit in the ICO scheme. The model is trained over $200$ epochs with a learning rate $0.01$ and batches of size $32$. We use a total of $K = 200$ training pairs in the dataset $\mathcal{D} = \{x_k,y_k\}_{k=1}^K$. We use a simulator and we compute the gradients with standard backpropagation. Results are averaged over five independent trials. 

%For the benchmark scheme, we use grid search to find the optimal hyperparameters.

\subsection{Benchmarks}\label{sec:bench}
As a benchmark, we consider the scheme described in Sec.~\ref{Sec:ICO}, using the same architectures  and hyperparameters for the ansatz as described in Sec.~\ref{sec:arch}. Thereby, in the benchmark scheme sensing and computation are carried out using a shared quantum state that performs state observation and then computes a function of the state. Here, sensing and computation follow a strict order, with the former always preceding  the later to test the impact of ICO. 

\subsection{Results}\label{sec:res}
We quantify the performance of the proposed scheme and the benchmark in terms of the accrued training and testing loss. The training loss is obtained using \eqref{mse}, and the testing loss is computed using $50$ samples of unobserved pairs of data samples $x_k$ and corresponding targets $y_k$.

We show the accrued training and testing loss in Fig~\ref{fig:entropy_Tr}. As it can be seen, the ICO scheme is able to learn the considered task and achieve a small training loss. In addition, it can generalise well to new data as evidenced by the small testing loss. We also compare the ICO scheme with the benchmark which follows a linear order of sensing then learning. We observe that the ICO scheme converges faster, requiring a smaller number of steps to minimise the loss function.

\section{Discussion and Conclusion}
The ideas presented in this paper do not in any way imply that classical agents, e.g., ChatGPT can answer a query before the prompter supplies one. This work only shows that, at the fundamental level, the tasks of information acquisition and learning can be manipulated in ways that transcend linear causality. Specifically, we proposed a framework for integrated sensing and computation, whereby both tasks are carried out using the same quantum state, and we investigated whether they can be carried out with ICO. 
%Under the ICO operation, the agent effectively experiences a superposition of orders, one where it performs state observation and then processes information, and another where the agent computes and then observes the state. 
We provided experimental results and we showed that the proposed scheme can achieve small training and testing losses on a representative task in magnetic based navigation. 
Whilst this investigation is primarily conceptual, these results show that agents that can integrate sensing and learning can become crucial parts of future quantum technologies, such as \cite{popovski20251q}.

\subsection{Limitations and Future Directions}
%\subsection{Future Directions}
The ICO operation is contingent upon the ability to implement the quantum SWITCH. Although the SWITCH has been experimentally validated \cite{goswami2020experiments}, the practical stability of SWITCH-like operations is currently debated \cite{molitor2024quantum} and extensions of this work should include this consideration. In addition, practical quantum systems are subject to noise and decoherence which would affect the performance of this scheme. Similarly, alternative parameterized circuits $U_{\theta}$, such as symmetry-preserving models could further improve performance and even simplify circuit complexity \cite{nikoloska2023time}. Techniques for robust training, such as advanced optimisers or regularization to avoid barren plateaus, are also worth exploring. Finally, the limits of an ICO operation should be investigated. Whilst the benefits of ICO in communication, the foundational form of information processing, have been formally established, it is important to theoretically characterise the performance of ICO in integrated sensing and computation to rigorously assess the role of causal order.

%Limits of ... when is it good and bad ...

\bibliographystyle{IEEEtran}
\bibliography{litdab}

\begin{thebibliography}{10}
\providecommand{\url}[1]{#1}
\csname url@samestyle\endcsname
\providecommand{\newblock}{\relax}
\providecommand{\bibinfo}[2]{#2}
\providecommand{\BIBentrySTDinterwordspacing}{\spaceskip=0pt\relax}
\providecommand{\BIBentryALTinterwordstretchfactor}{4}
\providecommand{\BIBentryALTinterwordspacing}{\spaceskip=\fontdimen2\font plus
\BIBentryALTinterwordstretchfactor\fontdimen3\font minus \fontdimen4\font\relax}
\providecommand{\BIBforeignlanguage}[2]{{%
\expandafter\ifx\csname l@#1\endcsname\relax
\typeout{** WARNING: IEEEtran.bst: No hyphenation pattern has been}%
\typeout{** loaded for the language `#1'. Using the pattern for}%
\typeout{** the default language instead.}%
\else
\language=\csname l@#1\endcsname
\fi
#2}}
\providecommand{\BIBdecl}{\relax}
\BIBdecl

\bibitem{sutton1998reinforcement}
R.~S. Sutton, A.~G. Barto \emph{et~al.}, \emph{Reinforcement learning: An introduction}.\hskip 1em plus 0.5em minus 0.4em\relax MIT press Cambridge, 1998, vol.~1, no.~1.

\bibitem{weaver1953recent}
W.~Weaver, ``Recent contributions to the mathematical theory of communication,'' \emph{ETC: a review of general semantics}, pp. 261--281, 1953.

\bibitem{chiribella2009beyond}
G.~Chiribella, G.~D’Ariano, P.~Perinotti, and B.~Valiron, ``Beyond quantum computers,'' \emph{arXiv preprint arXiv:0912.0195}, 2009.

\bibitem{hardy2007towards}
L.~Hardy, ``Towards quantum gravity: a framework for probabilistic theories with non-fixed causal structure,'' \emph{Journal of Physics A: Mathematical and Theoretical}, vol.~40, no.~12, p. 3081, 2007.

\bibitem{brassard2005information}
G.~Brassard, ``Is information the key?'' \emph{Nature Physics}, vol.~1, no.~1, pp. 2--4, 2005.

\bibitem{fuchs2003quantum}
C.~A. Fuchs, ``Quantum mechanics as quantum information, mostly,'' \emph{Journal of Modern Optics}, vol.~50, no. 6-7, pp. 987--1023, 2003.

\bibitem{zhao2025quantum}
X.~Zhao, B.~Zhao, and G.~Chiribella, ``The quantum communication power of indefinite causal order,'' \emph{arXiv preprint arXiv:2510.08507}, 2025.

\bibitem{guerin2016exponential}
P.~A. Gu{\'e}rin, A.~Feix, M.~Ara{\'u}jo, and {\v{C}}.~Brukner, ``Exponential communication complexity advantage from quantum superposition of the direction of communication,'' \emph{arXiv preprint arXiv:1605.07372}, 2016.

\bibitem{wang2026quantum}
Y.-X. Wang, F.~Salvati, D.~R. Shukur, W.~F. Braasch~Jr, K.~Murch, and N.~Y. Halpern, ``Quantum metrology enhanced by effective time reversal,'' \emph{arXiv preprint arXiv:2601.20952}, 2026.

\bibitem{yin2023experimental}
P.~Yin, X.~Zhao, Y.~Yang, Y.~Guo, W.-H. Zhang, G.-C. Li, Y.-J. Han, B.-H. Liu, J.-S. Xu, G.~Chiribella \emph{et~al.}, ``Experimental super-heisenberg quantum metrology with indefinite gate order,'' \emph{Nature Physics}, vol.~19, no.~8, pp. 1122--1127, 2023.

\bibitem{quintino2019reversing}
M.~T. Quintino, Q.~Dong, A.~Shimbo, A.~Soeda, and M.~Murao, ``Reversing unknown quantum transformations: Universal quantum circuit for inverting general unitary operations,'' \emph{Physical review letters}, vol. 123, no.~21, p. 210502, 2019.

\bibitem{quintino2019probabilistic}
------, ``Probabilistic exact universal quantum circuits for transforming unitary operations,'' \emph{Physical Review A}, vol. 100, no.~6, p. 062339, 2019.

\bibitem{goodfellow2016deep}
I.~Goodfellow, Y.~Bengio, A.~Courville, and Y.~Bengio, \emph{Deep learning}.\hskip 1em plus 0.5em minus 0.4em\relax MIT press Cambridge, 2016, vol.~1, no.~2.

\bibitem{meyer2021variational}
J.~J. Meyer, J.~Borregaard, and J.~Eisert, ``A variational toolbox for quantum multi-parameter estimation,'' \emph{npj Quantum Information}, vol.~7, no.~1, p.~89, 2021.

\bibitem{maclellan2024end}
B.~MacLellan, P.~Roztocki, S.~Czischek, and R.~G. Melko, ``End-to-end variational quantum sensing,'' \emph{arXiv preprint arXiv:2403.02394}, 2024.

\bibitem{nikoloska2025dynamic}
I.~Nikoloska, H.~Joudeh, R.~van Sloun, and O.~Simeone, ``Dynamic estimation loss control in variational quantum sensing via online conformal inference,'' \emph{arXiv preprint arXiv:2505.23389}, 2025.

\bibitem{nikoloska2025adaptivebayesiansingleshotquantum}
I.~Nikoloska, R.~Van~Sloun, and O.~Simeone, ``Adaptive bayesian single-shot quantum sensing,'' \emph{arXiv preprint arXiv:2507.16477}, 2025.

\bibitem{nikoloska2025variational}
I.~Nikoloska and O.~Simeone, ``Variational quantum integrated sensing and communication,'' \emph{arXiv preprint arXiv:2511.16597}, 2025.

\bibitem{nikoloska2026machine}
I.~Nikoloska, ``Machine learning with quantum computers,'' in \emph{Artificial Intelligence and Intelligent Matter: Nanoscience, Soft Matter, Philosophy}.\hskip 1em plus 0.5em minus 0.4em\relax Springer, 2026, pp. 417--434.

\bibitem{chiribella2021indefinite}
G.~Chiribella, M.~Banik, S.~S. Bhattacharya, T.~Guha, M.~Alimuddin, A.~Roy, S.~Saha, S.~Agrawal, and G.~Kar, ``Indefinite causal order enables perfect quantum communication with zero capacity channels,'' \emph{New Journal of Physics}, vol.~23, no.~3, p. 033039, 2021.

\bibitem{akai2014development}
N.~Akai, S.~A. Rahok, K.~Inoue, and K.~Ozaki, ``Development of magnetic navigation method based on distributed control system using magnetic and geometric landmarks,'' \emph{ROBOMECH Journal}, vol.~1, no.~1, p.~21, 2014.

\bibitem{ataka2022magnetic}
A.~Ataka, H.-K. Lam, and K.~Althoefer, ``Magnetic-field-inspired navigation for robots in complex and unknown environments,'' \emph{Frontiers in Robotics and AI}, vol.~9, p. 834177, 2022.

\bibitem{aiello2013composite}
C.~D. Aiello, M.~Hirose, and P.~Cappellaro, ``Composite-pulse magnetometry with a solid-state quantum sensor,'' \emph{Nature communications}, vol.~4, no.~1, p. 1419, 2013.

\bibitem{maayani2019distributed}
S.~Maayani, C.~Foy, D.~Englund, and Y.~Fink, ``Distributed quantum fiber magnetometry,'' \emph{Laser \& Photonics Reviews}, vol.~13, no.~7, p. 1900075, 2019.

\bibitem{popovski20251q}
\BIBentryALTinterwordspacing
P.~Popovski, Čedomir Stefanović, B.~Soret, I.~Leyva-Mayorga, S.~R. Pandey, R.~B. Christensen, J.~K. Søndergaard, K.~S. Jensen, T.~G. Pedersen, A.~S. Cacciapuoti, and L.~Hanzo, ``1q: First-generation wireless systems integrating classical and quantum communication,'' \emph{arXiv preprint arXiv:2509.14731}, 2025. [Online]. Available: \url{https://arxiv.org/abs/2509.14731}
\BIBentrySTDinterwordspacing

\bibitem{goswami2020experiments}
K.~Goswami and J.~Romero, ``Experiments on quantum causality,'' \emph{AVS Quantum Science}, vol.~2, no.~3, 2020.

\bibitem{molitor2024quantum}
O.~A. Molitor, A.~H. Malavazi, R.~D. Baldij{\~a}o, A.~C. Orthey~Jr, I.~L. Paiva, and P.~R. Dieguez, ``Quantum switch instabilities with an open control,'' \emph{Communications Physics}, vol.~7, no.~1, p. 373, 2024.

\bibitem{nikoloska2023time}
I.~Nikoloska, O.~Simeone, L.~Banchi, and P.~Veli{\v{c}}kovi{\'c}, ``Time-warping invariant quantum recurrent neural networks via quantum-classical adaptive gating,'' \emph{Machine Learning: Science and Technology}, 2023.

\end{thebibliography}

\end{document}